\documentclass[12pt,preprint]{aastex}

\usepackage{color}
\usepackage{ulem}

\shorttitle{Polarimetric Imaging of the Disk around PDS~70}
\shortauthors{Hashimoto et al.}

\begin{document}
  \title{Polarimetric Imaging of Large Cavity Structures in the Pre-transitional Protoplanetary Disk around PDS~70: Observations of the disk\footnote{
    Based on data collected at the Subaru Telescope, which is operated by the National Astronomical Observatory of Japan.}}

  \author{
    J.~Hashimoto\altaffilmark{1}, 
    R.~Dong\altaffilmark{2}, 
    T.~Kudo\altaffilmark{3}, 
    M.~Honda\altaffilmark{4},
    M.~K.~McClure\altaffilmark{5}, 
    Z.~Zhu\altaffilmark{2},
    T.~Muto\altaffilmark{6}, 
    J.~Wisniewski\altaffilmark{7,8},
    L.~Abe\altaffilmark{9}, 
    W.~Brandner\altaffilmark{10}, 
    T.~Brandt\altaffilmark{2}, 
    J.~Carson\altaffilmark{10,11},
    S.~Egner\altaffilmark{3}, 
    M.~Feldt\altaffilmark{10}, 
    M.~Fukagawa\altaffilmark{12},
    M.~Goto\altaffilmark{13}, 
    C.~A.~Grady\altaffilmark{14,15}, 
    O.~Guyon\altaffilmark{3}, 
    Y.~Hayano\altaffilmark{3},
    M.~Hayashi\altaffilmark{1}, 
    S.~Hayashi\altaffilmark{3}, 
    T.~Henning\altaffilmark{10}, 
    K.~Hodapp\altaffilmark{16},
    M.~Ishii\altaffilmark{3},
    M.~Iye\altaffilmark{1}, 
    M.~Janson\altaffilmark{2},
    R.~Kandori\altaffilmark{1}, 
    G.~Knapp\altaffilmark{2}, 
    N.~Kusakabe\altaffilmark{1},
    M.~Kuzuhara\altaffilmark{17,1},
    J.~Kwon\altaffilmark{18,1},
    T.~Matsuo\altaffilmark{19},
    S.~Mayama\altaffilmark{20},
    M.~W.~McElwain\altaffilmark{14}, 
    S.~Miyama\altaffilmark{21}, 
    J.-I.~Morino\altaffilmark{1},
    A.~Moro-Martin\altaffilmark{22,2}, 
    T.~Nishimura\altaffilmark{3}, 
    T.-S.~Pyo\altaffilmark{3},
    G.~Serabyn\altaffilmark{23},
    T.~Suenaga\altaffilmark{18,1},
    H.~Suto\altaffilmark{1}, 
    R.~Suzuki\altaffilmark{1},
    Y.~Takahashi\altaffilmark{24,1},
    M.~Takami\altaffilmark{25},
    N.~Takato\altaffilmark{3}, 
    H.~Terada\altaffilmark{3}, 
    C.~Thalmann\altaffilmark{26}, 
    D.~Tomono\altaffilmark{3},
    E.~L.~Turner\altaffilmark{2,27}, 
    M.~Watanabe\altaffilmark{28}, 
    T.~Yamada\altaffilmark{29}, 
    H.~Takami\altaffilmark{3},
    T.~Usuda\altaffilmark{3},
    M.~Tamura\altaffilmark{1}
    }
  
  \altaffiltext{1}{National Astronomical Observatory of Japan, 2-21-1 Osawa, Mitaka, Tokyo 181-8588, 
    Japan; jun.hashimoto@nao.ac.jp}
  \altaffiltext{2}{Department of Astrophysical Sciences, Princeton University, NJ 08544, USA}
  \altaffiltext{3}{Subaru Telescope, 650 North A'ohoku Place, Hilo, HI 96720, USA}
  \altaffiltext{4}{Kanagawa University, 2946 Tsuchiya, Hiratsuka, Kanagawa 259-1293, Japan}
  \altaffiltext{5}{Department of Astronomy, The University of Michigan, 500 Church St., 830 Dennison Bldg., Ann Arbor, MI 48109, USA}
  \altaffiltext{6}{Division of Liberal Arts, Kogakuin University, 1-24-2, Nishi-Shinjuku, Shinjuku-ku, Tokyo, 163-8677, Japan}
  \altaffiltext{7}{University of Washington, Seattle, Washington, USA}
  \altaffiltext{8}{HL Dodge Department of Physics \& Astronomy, University of Oklahoma, Norman, OK 73019, USA}
  \altaffiltext{9}{Laboratoire Hippolyte Fizeau, UMR6525, Universite de Nice Sophia-Antipolis, 28, avenue Valrose, 06108 Nice Cedex 02, France}
  \altaffiltext{10}{Max Planck Institute for Astronomy, Heidelberg, Germany}
  \altaffiltext{11}{Department of Physics and Astronomy, College of Charleston, 58 Coming St., Charleston, SC29424, USA}
  \altaffiltext{12}{Osaka University, 1-1, Machikaneyama, Toyonaka, Osaka 560-0043, Japan}
  \altaffiltext{13}{Universit\"ats-Sternwarte M\"unchen Scheinerstr. 1, D-81679 Munich, Germany}
  \altaffiltext{14}{Exoplanets and Stellar Astrophysics Laboratory, Code 667, Goddard Space Flight Center, Greenbelt, MD 20771 USA}
  \altaffiltext{15}{Eureka Scientific, 2452 Delmer, Suite 100, Oakland CA 96002, USA}
  \altaffiltext{16}{University of Hawaii,640 North A'ohoku Place, Hilo, HI 96720, USA}
  \altaffiltext{17}{Department of Earth and Planetary Science, University of Tokyo, 7-3-1 Hongo, Tokyo 113-0033, Japan}
  \altaffiltext{18}{Department of Astronomical Science, Graduate University for Advanced Studies (Sokendai), Tokyo 181-8588, Japan}
  \altaffiltext{19}{Department of Astronomy, Kyoto University, Kita-shirakawa-Oiwake-cho, Sakyo-ku, Kyoto 606-8502, Japan}
  \altaffiltext{20}{The Graduate University for Advanced Studies, Shonan International Village, Hayama-cho, Miura-gun, Kanagawa 240-0193, Japan}
  \altaffiltext{21}{Hiroshima University, 1-3-2, Kagamiyama, Higashi-Hiroshima 739-8511, Japan}
  \altaffiltext{22}{Department of Astrophysics, CAB - CSIC/INTA, 28850 Torrej'on de Ardoz, Madrid, Spain}
  \altaffiltext{23}{Jet Propulsion Laboratory, California Institute of Technology, 4800 Oak Grove Drive, Pasadena, CA 91109, USA}
  \altaffiltext{24}{Department of Astronomy, University of Tokyo, Tokyo, Japan}
  \altaffiltext{25}{Institute of Astronomy and Astrophysics, Academia Sinica, P.O. Box 23-141, Taipei 10617, Taiwan}
  \altaffiltext{26}{Anton Pannekoek Astronomical Institute, University of Amsterdam, Amsterdam, The Netherlands}
  \altaffiltext{27}{Kavli Institute for the Physics and Mathematics of the Universe, The University of Tokyo, Kashiwa 227-8568, Japan}
  \altaffiltext{28}{Department of Cosmosciences, Hokkaido University, Sapporo 060-0810, Japan}
  \altaffiltext{29}{Astronomical Institute, Tohoku University, Aoba, Sendai 980-8578, Japan}

  \begin{abstract}    
    We present high resolution $H$-band polarized intensity ($PI$; FWHM = $0.''1$: 14 AU) 
    and $L'$-band imaging data (FWHM $= 0.''11$: 15 AU)
    of the circumstellar disk around the weak-lined T Tauri star PDS~70 in Centaurus
    at a radial distance of 28 AU ($0.''2$) up to 210 AU (1.$''$5).
    In both images, a giant inner gap is clearly resolved for the first time, and 
    the radius of the gap is $\sim$70 AU. 
    Our data show that the geometric center of the disk shifts by $\sim$6 AU toward the minor axis.
    We confirm that the brown dwarf companion candidate to the north of PDS~70
    is a background star based on its proper motion.
    As a result of SED fitting by Monte Carlo radiative transfer modeling,
    we infer the existence of an optically thick inner disk at a few AU.  Combining our observations and modeling,
    we classify the disk of PDS~70 as a pre-transitional disk.
    Furthermore, based on the analysis of $L'$-band imaging data, we put 
    an upper limit mass of companions at $\sim$30 to $\sim$50$M_{\rm J}$ within the gap. 
   Taking account of the presence of the large and sharp
    gap, we suggest that
    the gap could be formed by dynamical interactions of sub-stellar companions or 
   multiple unseen giant planets in the gap. 
  \end{abstract}

  \section{Introduction}\label{intro}
  Protoplanetary disks are believed to be the birthplaces of planets \citep[e.g.,][]{haya85}; hence, 
  understanding the evolution of these disks guides our understanding of the process of planet formation.
  Disks which have substantial infrared excesses but reduced fluxes at 
  wavelengths $\lesssim$20 $\mu$m, i.e., transitional disks \citep{stro89}, could be related to the early phases of 
  planet formation \citep[see a recent review of][]{will11} and are therefore 
  particularly important for understanding how, where, and when planets form.
  For many transitional disks, partial inner holes or partial gaps have been directly 
  resolved by interferometry at (sub)millimeter wavelengths 
  \citep[e.g.,][]{piet06,andr11} and imaging at near-infrared wavelengths
  \citep{fuka06,thal10,hash11}. 
  Numerous mechanisms have been proposed to explain the clearing of gaps in transitional disks, including
  grain growth \citep[e.g.,][]{dull05}, photoevaporation \citep[e.g.,][]{clar01}, 
  and gravitational interactions with orbiting planets 
  \citep[e.g.,][]{papa07,zhu11}. Two possible methods
  to distinguish the disk-planet interactions from other aforementioned proposed gap-clearing
  mechanisms 
  could be the detection of (1) a planetary companion in the inner hole/gap region \citep[e.g.,][]{krau12}
  or (2) a ring-like gap between optically thick inner and outer disks 
  \citep[i.e., pre-transitional disk;][]{espa07} because
  dynamical formation of wide gaps could be the only surviving mechanism for wide gapped disks
  \citep[e.g.,][]{papa07,zhu11}
  
  One good candidate to investigate the inner hole/gap region at tens AU in the disk in pre-transitional disks 
  is the weak-lined T Tauri star PDS~70
  \citep[K5 type; 0.82$M_{\odot}$; $<$10 Myr;][]{greg02,riau06,metc04}. 
  A scattered light disk with a radius at 14 to 140 AU was detected by $Ks$-band imaging \citep{riau06}.
  The possible presence of inner and outer disks with different temperatures
  were suggested by \citet{metc04} and \citet{riau06}, which may imply that PDS~70 is a pre-transitional disk object.
  In this $Letter$, we present high resolution imaging of PDS~70 with Subaru/HiCIAO and Gemini/NICI.

  \section{Observations \& Data Reduction}\label{obs}
  \subsection{$H$-band Polarimetry with Subaru/HiCIAO}\label{obs_pol}
  Observations of PDS~70 were conducted with HiCIAO \citep{tamu06}
  on the Subaru telescope in polarized differential imaging (PDI) mode,
  combined with angular differential imaging (ADI) mode \citep{maro06},
  on 2012 February 28 UT. $H$-band linear polarization images were taken under 
  the program SEEDS \citep[Strategic Explorations of Exoplanets and Disks with Subaru;][]{tamu09}. 
  In PDI+ADI, we employed a double wollaston prism to split incident light into four images, 
  each has 5$''$ by 5$''$ field of view with a pixel scale of 9.5 mas/pixel,
  to make the saturated radius as small as possible. 

  Polarization was measured by rotating the half waveplate to four angular positions 
  (in the order of 0$^{\circ}$, 45$^{\circ}$, 22.5$^{\circ}$, 
  and 67.5$^{\circ}$). We obtained 22 full waveplate rotation cycles, taking 
  a 15-s exposure per waveplate position. The 
  total integration time of the $PI$ image was 660 s. 
  Even with a large airmass of $\sim$2 during observations of PDS~70,
  the adaptive optics system \citep[AO188;][]{haya04} provided a stellar PSF of PDS~70 with 
  FWHM of 0.$''$1 in the $H$ band.
  The angle of the total field rotation was $\sim$13$^{\circ}$.
 
\begin{figure}
    \epsscale{1}
    \plotone{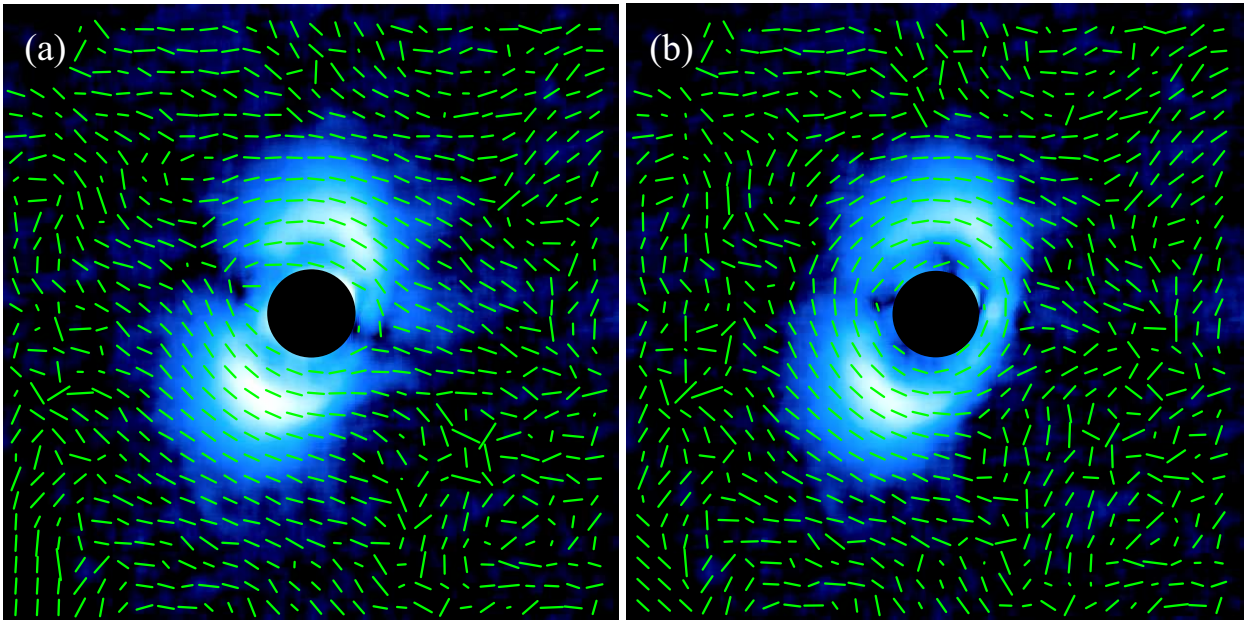}
    \caption{$H$-band polarization vectors of PDS~70 are superposed on the $PI$ image 
      with a software mask with 0.$''$4 diameter before subtracting polarized halo 
      (a) and after subtraction (b). The plotted vectors are
      binned with spatial resolution. The field of view (FOV) is 3.$''$0 $\times$ 3.$''$0. 
      All plotted vectors' lengths are arbitrary for the presentation purpose. 
    \label{f1}}
  \end{figure}

  The polarimetric data were reduced with the same procedure as for \citet{hash11} using IRAF\footnote{
  IRAF is distributed by the National Optical Astronomy Observatories, which are operated
  by the Association of Universities for Research in Astronomy,
  Inc., under cooperative agreement with the National Science Foundation.}.   
  Since the Strehl ratio of the stellar PSF of PDS~70 was $\sim$0.26, a stellar halo remained
  in the image, 
  i.e, the PSF convolved by seeing was not perfectly corrected by the AO and partially remains.
  This halo appears to be polarized along the minor axis of disks 
  since the polarization of forward and back scattering is smaller 
  due to the deviation from 90$^{\circ}$ in the scattering angle, 
  and then the stellar halo has a significant net-polarization. 
  In other words, since the flux of a PSF convolved by seeing
  contains contributions from both the central star and disk, the seeing-PSF might be
  polarized along the minor axis of a disk.
  To correct for this $``$polarized halo$"$, we first derived the net-polarizarion of PDS~70 with aperture polarimetry. 
  We then made a model $``$polarized halo$"$ of PDS~70 by multiplying the added image of $o$- and $e$-rays by the 
  derived polarization of $P = 0.503\% \pm 0.001\%$ with $\theta = 66.32^{\circ} \pm 0.06^{\circ}$. 
  The errors were calculated with the photometric errors, and are thus a lower limit. Finally,
  we subtracted the model $``$polarized halo$"$ and obtained final Stokes $Q$ and $U$ images.
  Figure~\ref{f1} demonstrates the subtraction of the $``$polarized halo$"$.

  \subsection{$L'$-band imaging with Gemini/NICI}\label{loci}
  We carried out ADI observations of PDS~70
  in the $L'$ band on 31 March 2012, using the Near-Infrared Coronagraphic
  Imager (NICI) and the 85-element AO system \citep{chun08} mounted on
  the 8.1 m Gemini South telescope at Cerro Pachon, Chili. NICI utilizes a
  1024 $\times$ 1024 ALADDIN InSb array with an plate scale of 18 mas/pix and field
  of view of 18 $\times$ 18 arcsec.
  
  We obtained 145 frames for PDS~70 with 0.76-s $\times$ 25 coadd for $L'$ band.
  Although the sky was clear for the full night, the seeing was variable,
  therefore a subset of 95 frames were
  combined to produce the final image. The total
  exposure time was 1805 sec and the spatial resolution was achieved 0.$''$11. 
  The angle of the total field rotation is $\sim$100$^{\circ}$.

  The data reduction was performed with the LOCI algorithm \citep{maro06,lafr07}
  in annular regions of   1000 $\times$ FWHM ($N_{\rm A} =$ 1000) with reference
  images selected from frames with at least 0.5 $\times$ FWHM  field rotation 
  ($N_{\delta}=$ 0.5). A large optimization area diminishes the impact of the disk
  on the optimization process \citep{buen10}.
  The ratio of radial and azimuthal width of the optimization area ($g$) is 1.

  \section{Results \& Discussion}
  
  \subsection{$H$-band Polarimetry and $L'$-band imaging} 
\begin{figure}
    \epsscale{1}
    \plotone{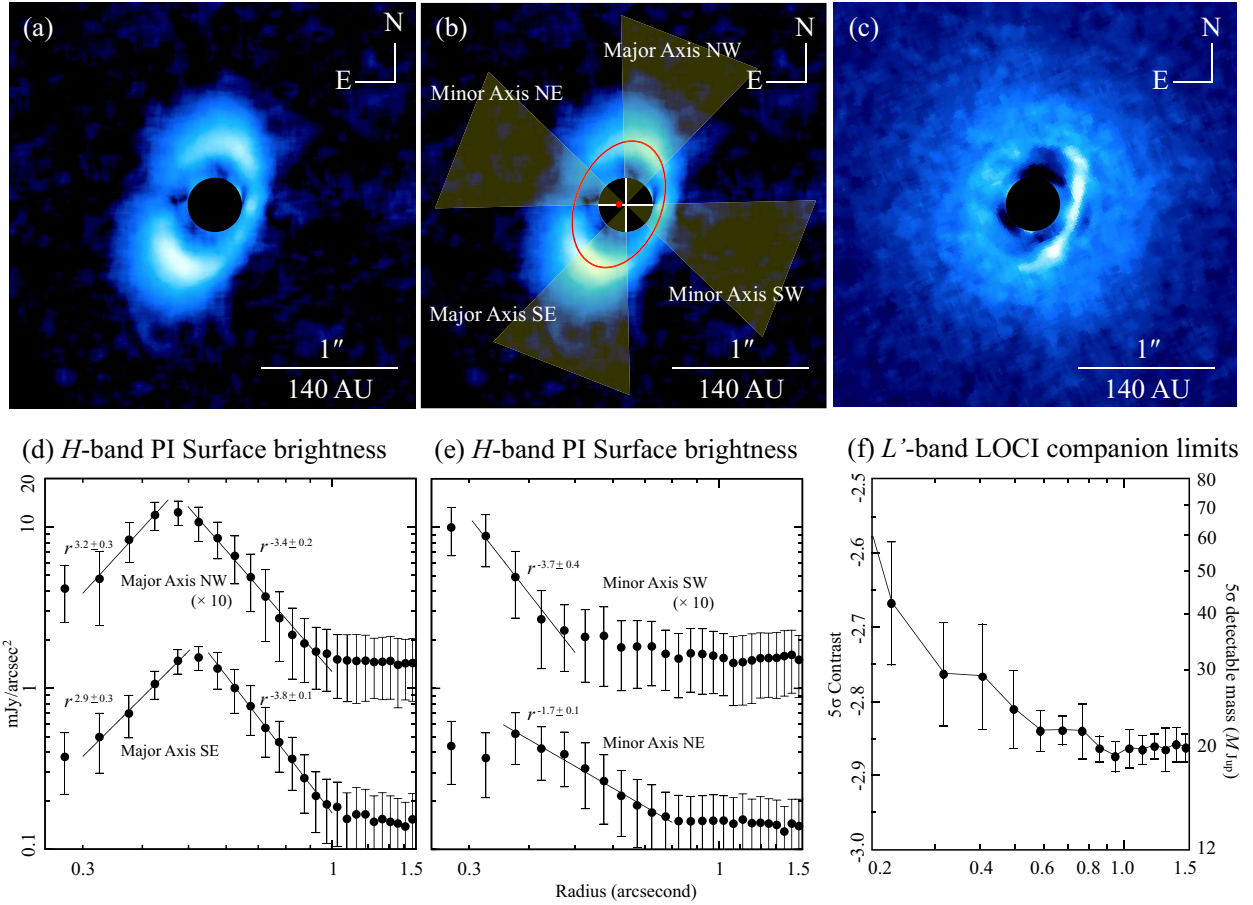}
    \caption{(a): $H$-band $PI$ image of PDS~70 with a software mask with 0.$''$4 diameter.
      (b): same with (a), but its features. The solid ellipse indicate the ring-like disk.
      The filled circle represent the geometric center of the disk. (c): $L'$-band LOCI image
      of PDS~70 with a software mask with 0.$''$4 diameter. 
      The parameters in LOCI reductions are described in sec.~\ref{loci}. The FOV of three images
      are 3.$''$0 $\times$ 3.$''$0 with a convolution of a spatial resolution.
      (d) and (e): Radial profiles at yellow hatched regions 
      of minor and major axes in (b). The values of the profile at northwest and 
      southwest are multiplied by ten for the presentation purpose. 
      (f): Detectable mass at 5$\sigma$ based on the $L'$-band LOCI image.
      The LOCI parameters are same with those described in sec.~\ref{loci}, 
      but the optimization area is 250~$\times$~FWHM ($N_{\rm A} =$ 250).
      \label{f2}}
  \end{figure}

  Figure~\ref{f2} shows $H$-band PI images and the $L'$-band LOCI image of 
  PDS~70 assuming a distance of 140 pc, along with radial surface brightness profiles.
  We find a clear elliptical ring in the $H$-band, 
  which has not been reported in previous high-resolution imaging \citep{riau06}.
  A partial elliptical disk is observed in the $L'$-band, 
  due to the inevitable loss of flux in the process of LOCI; hence, 
  we derive radial profiles of suface brightness based on the $H$-band PI image only
  and companion mass limits from the $L'$-band LOCI image only.
  
  We consider that the ellipse shape is due to the system's inclination,
  and show the results of fitting an ellipse to these data in Table~\ref{table1}. 
  The position angle of the major axis and the inclination 
  of the disk are similar with those of $K_{\rm s}$-band imaging 
  \citep[PA $\approx155^{\circ}$ and $i\approx62^{\circ}$;][]{riau06}.
  We measured an offset of 44 $\pm$ 3 mas ($\sim$6 AU) at PA=87.9$^{\circ}$
  between the geometric center of the disk and the central star. 
  The positional accuracy of the central star is 1.5 mas (0.2 AU).
  The direction of this offset is roughly consistent with that of the minor axis of 68.6$^{\circ}$, and the sign of this offset indicates that the southwest side is inclined toward us (i.e. the near side) (see the model image in \citealt{dong12b}). This geometry is also consistent with the facts that (1) the northeast side of the disk is wider due to the back illumination of the wall \citep{thal10}, and (2) the southwest side is brighter than the northwest side due to forward scattering \citep{fuka06}.

  
  Assuming the cavity edges correspond to the peak PI of the disk, 
  a radius of the cavity is measured as $\sim$70 AU.
  The outer radius of the disk is measured to be $\sim$140 AU in figure~\ref{f2}(d), and corresponds to 
  the location at which our sensitivity is no longer sufficient to detect extended emission.

  A single power-law fit was performed to the radial profiles along the minor and major axes 
  (figure~\ref{f2}d and e). 
  Our results of $\sim$ -3.6 and $\sim$ -1.7 are different from $\sim$ -2.8 of \citet{riau06}.

  We also found a flux deficit of $PI$ in the direction of the minor axis. 
  Since the observed scattering angle at the minor axis
  deviates from 90$^{\circ}$, the polarization fraction along the minor axis is lower.
  The $PI$ at the minor axis is therefore lower than at the major axis, and
  such hole-like structures are similar to those discussed in \citet{perr09}.
  
  We checked the proper motion of the companion cadidate
  to the north of PDS~70 reported by \citet{riau06}.  PDS~70 has a proper motion of 
  $(\mu_{\alpha}{\rm cos} \ \delta, \mu_{\delta}) = (-24.7 \pm 11.4, -13.3 \pm 11.4)$ mas/yr
  \citep{rose10}; the separation between PDS~70 and the companion cadidate should increase
  if the companion cadidate is a background star. 
  Since the separation in HiCIAO and NICI images are 
  324.13 $\pm$ 0.15 AU 
  and 324.44 $\pm$ 0.10 AU, respectively,
  the separation in \citet{riau06} is inferred to be 
  $309 \pm 12$ AU.
  Our estimation has a good agreement with the actual observed separation
  of 301.75 $\pm$ 0.06 AU in \citet{riau06}, and therefore,
  we concluded that the companion cadidate is a background star.

  \subsection{Detectable planetary-mass companions}
  Since the follow-up $L'$-band observations with Gemini/NICI failed to detect 
  any significant signals of point-like sources in the gap 
  we put constraints of upper limits for companion(s).
  Figure~\ref{f2}(f) shows the detectable masses of companions at 5$\sigma$.
  The LOCI parameter of the optimization area
  is 250 $\times$ FWHM, which is different from that described in sec.~\ref{loci}. 
  For that, we first applied a median filter with 0.$''$11
  width to the image, and then calculated the standard deviation as a noise level
  in concentric annuli along the major axis of the disk. The mass
  was calculated by assuming the COND evolutionary model \citep{bara03}, a distance
  of 140 pc, and an age of 10 Myr \citep{metc04}. We took into account
  the flux loss due to the partial self-subtraction by testing how point sources
  are affected 
  by LOCI. 
  The detectable mass limit is
  tens $M_{\rm J}$,
  therefore, stellar companions down to masses associated with massive brown dwarfs are excluded within the gap.

  \subsection{Modeling of spectral energy distribution (SED)}\label{model} 
  Although the SED fitting for PDS~70 has been performed in previous studies
  \citep{metc04,riau06}, both the availability of new archival photometric data (table~\ref{table2}) 
  and our imaging results motivates us to revisit the SED of the system using 
  Monte Carlo radiative transfer (MCRT). 
  Note that though PDS~70 has been observed by Spitzer IRS, the object was
  mispointed by $\sim$2.3$''$ with components along both the spatial and
  dispersion axes, therefore, the IRS spectrum of PDS~70 is not used in this work.

  {\bf Setup for modeling.} The method for our MCRT simulations is described in \citet{dong12a} and 
  \citet[in prep.]{whit12}. In a subsequent paper II \citep[in prep.]{dong12b},
  we will perform a detailed radiative transfer modeling
  of both the SED and the SEEDS imagery, and present a fiducial disk+cavity
  model which reproduces both observations well. We will also explore the
  parameter space around the fiducial model in paper II, and provide a
  full discussion on the constraints on the various model parameters there.
  In this letter, we only briefly describe the fiducial model and its
  resulting SED. We note that this model is not a {\it best} fitting model in an absolute sense, 
  since a full $\chi^2$ fitting of the observations with
  all the free parameters in a protoplanetary disk is essentially impossible
  \citep{math12}. However the constraints on many parameters such as the cavity size, depletion, and the surface
  density of the inner disk are reasonably tight, as will be shown in paper II.

  Our model contains a cavity 70 AU in radius. The surface density both inside and outside the cavity decreases with radius as 
  $\Sigma\propto \frac{R_c}{R}e^{-R/R_c}$, where $R_c=50$ AU, while $\Sigma$ inside the cavity is reduced to 
  $\delta\times$ the extrapolated value from the outer disk, with $\delta$ being the depletion factor.
  The temperature structure of the disk is determined from the radiative transfer calculations.
  The inner edge of the disk is self-consistently determined at the dust  
  sublimation temperature ($\sim1600$ K). We ignore accretion in the model,
  as suggested by its nature of being a weak line T Tauri star \citep{greg02}. 
  We use a pre-main sequence star of spectral type K5, radius 1.39 $R_\odot$, mass 0.82 $M_\odot$,
  and temperature 4500~K for the central source, as suggested by \citet{greg02} 
  and \citet{riau06}. The disk has a gaussian density distribution in the vertical direction, with scale heights $h$ as input parameters. 
  Two disk components are included: a thick disk with small grains (sub-micron-sized), 
  representing the pristine ISM-like dust; and a thin disk with large grains (up to $\sim$mm-sized) 
  and 20\% of the scale height of the small grains, as the result of grain growth and settling 
  \citep{dull04a,dull04b,dull05}. 
  Most of the dust mass (0.967) is in the settled disk, and the total dust-to-gas ratio 
  is assumed to be 1/100.   
  The SED is produced assuming a disk inclination angle $50^\circ$.
  The other parameters are summarized in table~\ref{table1}.

\begin{figure}
    \epsscale{1}
    \plotone{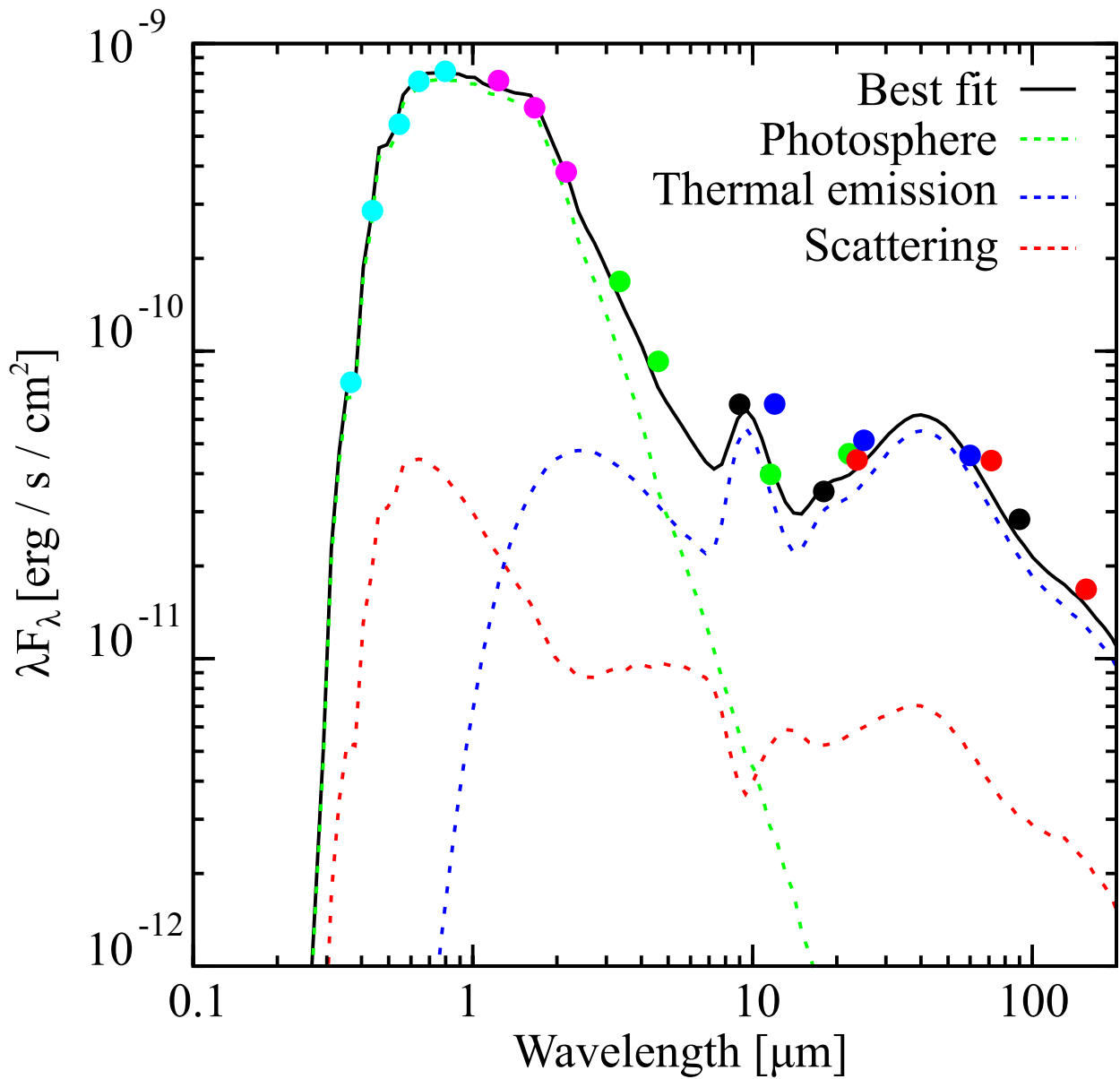}
    \caption{Pre-transitional disk model of PDS~70. Filled circles represent
	archival photometry (see table~\ref{table2} for photometry data).
	The solid black line is the best-fit model
	with a gap of $\sim$70 AU (see table~\ref{table1} for model parameters).
	Separate model components are as follows: stellar photosphere (green dotted line),
	thermal emission (blue dotted line), and scattered light (red dotted line). 
    \label{f3}}
  \end{figure}

  {\bf SED fitting.} Figure~\ref{f3} shows the good agreement between our model SED and available photometric data.
  As we will show in paper II, the thermal emission from the cavity wall 
  at $\sim$70 AU peaks at $\sim40\mu$m. The wall emission is a major signature of (pre-)transitional disk SED 
  \citep[e.g.,][]{espa07}.

  The surface density of small dust is $\sim0.001$g cm$^{-2}$ at 0.1 AU, and the opacity of small dust is 
  $\sim10^4$cm$^2$ g$^{-1}$ at $\sim1\mu$m (roughly the peak of the stellar radiation). This makes the 
  inner disk optically thick in the vertical direction. Since we assume a surface density profile 
  decreasing with increasing radius, when moving out the cavity gradually becomes vertically 
  optically thin, and the transition happens at $\sim1$ AU if only taking into account the small dust 
  (the existence of big dust inside the cavity is poorly determined from SED and scattered light image). 
  As we will discuss in detail in paper II, this optically thick inner disk is needed to explain
  the $\sim$2-40 $\mu$m SED, as models with optically thin
  inner disk fail to reproduce the SED. Following the convention in the literature 
  \citep{espa07}, we classify PDS~70 as a pre-transitional disk object.
  
  \subsection{Origin of the gap}
    
  Grain growth is one mechanism which can potentially form disk gaps \citep[e.g.,][]{birn12}. 
  It is capable of reproducing the SED of transitional disks 
  but not observed millimeter-wavelength images.
  However, the sharpness of PDS~70's $H$-band gap edge suggests grain growth is unlikely 
  to be the primary reason, since the models generally predict smooth gap edge features \citep[e.g.,][]{birn12}.

  Photoevaporation is another possible mechanism.  
  A photoevaporative wind can prevent outer disk material from feeding 
  the inner disk; without this supply, the inner disk material will rapidly accrete onto the central star, creating 
  a cavity inside out \citep{clar01}. Although a pre-transitional disk-like structure
  could be produced for a very short period \citep[see figure~1 in][]{alex06},
  it is quite unlikely to observe such a snap-shot by coincidence, so that it may be difficult to explain
  the existence of the optically thick inner disk inferred from the SED.
 
  Dynamical interactions with (sub)stellar companions 
  \citep[e.g.,][]{arty94}
  or orbiting planets \citep[e.g.,][]{papa07,zhu11} are also potential mechanisms,
  as several binary systems \citep[e.g., CoKu Tau 4;][]{irel08} are known to have a cavity
  in their circumbinary disk, and simulations show that multiple planets can open wide gaps \citep{zhu11},
  Furthermore, the survival of an optically thick inner 
  disk within a few AU in \citet{zhu11} mimics the inner disk of PDS~70.
  Our $L'$-band observation has only ruled out companions with a mass
  of $\gtrsim$50$M_{\rm J}$ in the gap region. 
  Therefore, dynamical interactions with low mass
  brown dawarfs or giant planets may be the origin of the gap.
  To put a futher constraint of an upper limit mass, future observations should be pursued, such as the aperture-masking interferometric
  observations \citep[e.g.,][]{naka89}, which has a better contrast 
  than LOCI imaging. Since PDS~70 harbors
  the pre-transitional disk with one of the largest gaps, this object may be one of the
  best candidate for future high-contrast planet imagers.

  \bigskip  
  We are grateful to an anonymous referee for providing
  many useful comments leading to an improved paper.
  We appreciate support from the Subaru Telescope and the Gemini South Telescope staff,
  especially from Jennie Berghuis and Dr.~Tom Hayward.
  This work is partly supported by a Grant-in-Aid for Science Research in
  a Priority Area from MEXT, by the Mitsubishi Foundation, 
  and by the U.S. National Science Foundation under Awards No. 1009203 and 1009314.

  \clearpage
    
  \begin{table}
    \begin{center}
      \caption{The results of the ellipse and the SED fitting for the disk of PDS~70\label{table1}}
      \begin{tabular}{cc} 
        \tableline
        \tableline
	\multicolumn{2}{c}{Ellipse fitting\tablenotemark{1}} \\
	\tableline
        Diameter of the major axis\tablenotemark{2} (AU)               & 137.0 $\pm$ 1.1 \\
        Diameter of the minor axis\tablenotemark{2} (AU)               &  88.6 $\pm$ 0.5 \\
        Position angle of the major axis ($^{\circ}$)  & 158.6 $\pm$ 0.7  \\
        Inclination by ellipse fitting\tablenotemark{3} ($^{\circ}$)      &  49.7 $\pm$ 0.3  \\
        Geometric center\tablenotemark{2,4} (AU : AU ) & ($6.1$ $\pm$ 0.3 : 0.2 $\pm$ 0.4) \\
	\tableline
	\multicolumn{2}{c}{SED fitting\tablenotemark{5}} \\
	\tableline
	$M_{\rm disk}$ & 0.003 \\
	$f$ & 0.967 \\
	$h^{\rm o}_{\rm b}$(100AU) & 2         \\
	$b^{\rm o}_{\rm b}$ & 1.2       \\
	$h^{\rm c}_{\rm b}$(100AU) & 2         \\
	$b^{\rm c}_{\rm b}$ & 1.2       \\
	$\delta_{\rm cav,b}$ & $10^{-3}$ \\
	$h^{\rm o}_{\rm s}$(100AU) & 10        \\
	$b^{\rm o}_{\rm s}$ & 1.2       \\
	$h^{\rm c}_{\rm s}$(100AU) & 10        \\
	$b^{\rm c}_{\rm s}$ & 1.2       \\
	$\delta_{\rm cav,s}$ & $10^{-3}$ \\
	\tableline
      \end{tabular}
      \tablenotetext{1}{
        In the ellipse fitting for the disk, the peak positions were 
	first directly determined by the 
        radial profile at position angles every 5$^{\circ}$ the disk. 
        We then conducted an ellipse fit using an implementation of the nonlinear 
	least-squares Marquardt-Levenberg algorithm with five 
        free parameters of lengths for the major and minor axes, position angle, and central positions.
      }
      \tablenotetext{2}{
        We assume that the distance of PDS~70 is 140 pc.
      }
      \tablenotetext{3}{
        Derived from the ratio of the major and minor axes.
      }
      \tablenotetext{4}{
        Central position (0, 0) is corresponding to the stellar position.
      }
      \tablenotetext{5}{
	Row (1): Total mass of the disk (assuming a dust-to-gas-ratio 100). 
	Row (2): Mass fraction of big dust in total dust. Rows (3), (5), (8), and (10): 
	Scale height $h$ at 100 AU. 
	Subscripts $b$ and $s$ indicate big dust and small dust, respectively.
        Superscripts $o$ and $c$ indicate outer disk and cavity, respectively. 
	Rows (4), (6), (9), and (11): Power index $b$ in $h \propto R^{b}$ in various disk components. 
	Row (7) and (12): Depletion factor of the big and small dust disk.
      }
    \end{center}
  \end{table}
  
  \clearpage
  
  \begin{table}
    \begin{center}
      \caption{Archival photometry data for PDS~70\label{table2}}
      \begin{tabular}{lccc} 
        \tableline
        \tableline
	Wavelength                         & $F_{\nu}$ (mJy)    & Note          & Plotted color\\
	                                   &                    &               & in fig.~\ref{f3} \\
	\tableline
	$U$\tablenotemark{a,b}             &     9.7            & \citet{greg92} & cyan    \\
	$B$\tablenotemark{a,b}             &    41.8            & \citet{greg92} & cyan    \\
	$V$\tablenotemark{a,b}             &    99.2            & \citet{greg92} & cyan    \\
	$R$\tablenotemark{a,b}             &   160.9            & \citet{greg92} & cyan    \\
	$I$\tablenotemark{a,b}             &   216.3            & \citet{greg92} & cyan    \\
	2MASS ($J$)\tablenotemark{a,b}     &   311.9 $\pm$  6.9 & \citet{cutr03} & magenta \\
	2MASS ($H$)\tablenotemark{a,b}     &   342.7 $\pm$ 12.6 & \citet{cutr03} & magenta \\
	2MASS ($Ks$)\tablenotemark{a,b}    &   275.6 $\pm$  5.8 & \citet{cutr03} & magenta \\
	WISE (3.4 $\mu$m)\tablenotemark{b} &   188.1 $\pm$  4.0 & \citet{cutr12} & green   \\
	WISE (4.6 $\mu$m)\tablenotemark{b} &   142.0 $\pm$  2.6 & \citet{cutr12} & green   \\
	WISE (12 $\mu$m)\tablenotemark{b}  &   153.9 $\pm$  2.3 & \citet{cutr12} & green   \\
	WISE (22 $\mu$m)\tablenotemark{b}  &   341.8 $\pm$  0.7 & \citet{cutr12} & green   \\
	AKARI (9 $\mu$m)                   &   201.2 $\pm$ 25.8 & VizieR II/297  & black   \\
	AKARI (18 $\mu$m)                  &   209.8 $\pm$ 13.4 & VizieR II/297  & black   \\
	AKARI (90 $\mu$m)                  &   851.1 $\pm$ 62.6 & VizieR II/298  & black   \\
	IRAS (12 $\mu$m)                   &   251   $\pm$ 25.1 & \citet{mosh89} & blue    \\
	IRAS (25 $\mu$m)                   &   348   $\pm$ 27.8 & \citet{mosh89} & blue    \\
	IRAS (60 $\mu$m)                   &   884   $\pm$ 61.9 & \citet{mosh89} & blue    \\
	MIPS (24 $\mu$m)                   & 349.7\tablenotemark{c}   $\pm$ 7.0  & Spitzer Heritage Archive\tablenotemark{e} & red \\
	MIPS (70 $\mu$m)                   & 1049.9\tablenotemark{c}  $\pm$ 19.9 & Spitzer Heritage Archive\tablenotemark{e} & red \\
	MIPS (160 $\mu$m)                  & 873.0\tablenotemark{c,d} $\pm$ 36.2 & Spitzer Heritage Archive\tablenotemark{e} & red \\
	\tableline
      \end{tabular}
      \tablenotetext{a}{
        An extinction law was adopted from \citet{math90}.
      }
      \tablenotetext{b}{
        Absolute flux conversions in optical, 2MASS, WISE photometric data were adopted
          from \citet{bess98}, \citet{cohe03}, and \citet{jarr11}, respectively.
      }
      \tablenotetext{c}{
	Since photometric data were not available, we conducted aperture photometry
	for archival images.
      }
      \tablenotetext{d}{
        Since the sky value was not sufficiently measured due to a small field-of-view
	of available data, our photometry may be less reliable.
      }
      \tablenotetext{e}{
	This work is based in part on observations made with the Spitzer Space Telescope, obtained from the 
        NASA/ IPAC Infrared Science Archive, both of which are operated by the Jet Propulsion Laboratory, 
        California Institute of Technology under a contract with the National Aeronautics and Space Administration.
      }
    \end{center}
  \end{table}  

  \clearpage
  
  \clearpage

  \clearpage
  
\end{document}